\documentclass[9pt]{article}
\usepackage{amsfonts}
\usepackage{bbm}
\usepackage{latexsym}
\usepackage{graphicx}
\usepackage{amsfonts,amssymb}
\usepackage[T1]{fontenc}
\begin{document}
\title{\textbf{Analyzing Netizen's View and Reply Behaviors on the Forum}}
\author{Jiefei Yu$^{1}$, Yanqing Hu$^{1}$, Min Yu$^{2}$, Zengru Di$^{1}$\footnote{Author for correspondence: zdi@bnu.edu.cn}\\
\\\emph{ 1. Department of Systems Science, School of Management,}\\
\emph{Beijing Normal University, Beijing 100875, P.R.China}
\\\emph{2. Department of Information Technology and Management,}\\ \emph{ Beijing Normal University, Beijing 100875, P.R.China}}
\maketitle

\begin{abstract}

Quantitative understanding of human behaviors provides elementary
comprehension of the complexity of many human-initiated systems. In
this paper, we investigate the behavior of people on the $BBS$ forum
by the statistical analysis of the amounts of view and reply of
posts. According to our statistics, we find that the amounts of view
and reply of posts follow the power law distributions with different
power exponent. Furthermore, we discover that the amounts of view
and reply of posts have nonlinear relationship. They are related by
power function and show us straight line in log-log plot. Based on
the estimation of slope and intercept of the line, we can
characterize the behaviors quantitatively and know that people of
Chinese forum and those of foreign forum have different preference
towards replying to and viewing the posts. At last, we analyze the
burstiness and memory in replying time series. They show some
universal properties among different forum. All of them locate
themselves in the high-$B$, low-$M$ region.
\end{abstract}


\section{Introduction}

Human behavior, as an academic issue in science, has a history about
one century from Watson \cite{1}. As a joint interest of sociology,
psychology and economics, human behavior has been extensively
investigated during the last decades. A basic assumption embedded in
the previous analysis on human dynamics is that its temporal
statistics are uniform and stationary, which can be properly
described by a Poisson process. Accordingly, the interevent time
distribution should have an exponential tail \cite{2}. However,
through studying the distributions of interevent time of the human
behavior such as the people to send out or reply E-mail and surface
mail, Albert Barab\'asi discovered that these human behaviors
present obvious deviation to the Poisson process: long time silent
and short time high frequency eruption simultaneously present in
these human behaviors, and their distributions of interevent time
have power law fat tail \cite{3}.

From E-mail communications \cite{3,4} to surface mail communications
\cite{5,6,7} and short message communications \cite{8}, from
financial activities to library loans \cite{6}, from web browsing
\cite{9} to on-line movie watchings \cite{10}, and more and more
examples show that the distributions of interevent time of the human
behavior can not be described by a Poisson process but may use power
function to fit well. According to the previous studies about human
dynamics, we find that its methodology, excavating statistical laws
from the historical records of human activities, can apply to
analyze other issues. For instance, we can also use some of those
datasets to quantify the herd behavior of an individual, that is to
say, following the opinions of the majority of people in his/her
social surrounding in an irrational way.

With the development of information technology, web surfing has
become a part of people's daily life, including E-mail
communication, dealing with financial matters, reading news online,
browsing web, downloading software and so on. In the cyber world,
$BBS$ forum provides people a platform for mutual communication.
Those behaviors like ``lurking, spamming, sofast'', etc. are the
ways in which Netizens publicly express their ideas. The means of
expression is so unique for its dual role of ``openness'' and
``concealment''. Therefore the $BBS$ forum is loved by the majority
of Internet users.

On the $BBS$ forum, the new threads and Internet users are
increasing, and the Internet users can browse information and
express their own points. In the process of browsing information and
expressing their own points, the most common group is formed on the
$BBS$ forum, that is to say, browsing the posts of the same subject
will form a provisional group, different themes has different
groups. However, in the provisional group, some people simply browse
information, and others will publish their own opinions. Then there
are the amounts of view and reply on the $BBS$ forum. So what is the
difference between the different groups, and how are these
differences reflected are the interesting issues to be researched.
In our paper, we focus on the actions of Internet users through the
$BBS$ forum to solve these problems. The dates come from the $Sina$
Forum and $Chat Avenue$ Forum, the first one is the largest Chinese
community and the other one is the current leader in free online
chat rooms.

Our paper is organized as follows. In Sec.\ref{sect2}, we present
that the amounts of view and reply of posts which follow the power
law distributions and have different power exponents. In
Sec.\ref{sect3}, we address the nonlinearity relationship of the
amounts of view and reply of posts and we also calculate the fitted
line to characterize the forum. According to the fitted line, we
find that the differences between the different forums. In
Sec.\ref{sect4}, we analyze the burstiness and memory in replying
time series.

\section{The amounts of view and reply}\label{sect2}

Under computer program's help, we can carry on the statistics to the
amounts of view and reply about posts of different themes. Here we
collect the dates from $Sina$ Community Forum and $Chat Avenue$
Forum.

$Sina$ Community Forum is the world's largest Chinese community and
the Internet's most comprehensive well-known $BBS$, has a huge core
group of users, the theme of the plate covers culture, life,
society, current affairs, sports, entertainment and other areas
\cite{11}. We got the amounts of view and reply about posts of the
entertainment, sports, life, economics themes from $Sina$ Community
Forum.

Economic Commentary: From April 17, 2001 to November 11, 2008, there
were 32,480 posts. Pet Club: From December 1, 2005 to November 17,
2008, there were 39,950 posts. China Football: From June 14, 2001 to
July 13, 2008, there were 95,891 posts. Pastime Puzzles: From
November 19, 2004 to July 8, 2008, there were 49,346 posts.

\begin{figure}[th]
\begin {center}
\includegraphics[height=10cm,width=0.95\textwidth]{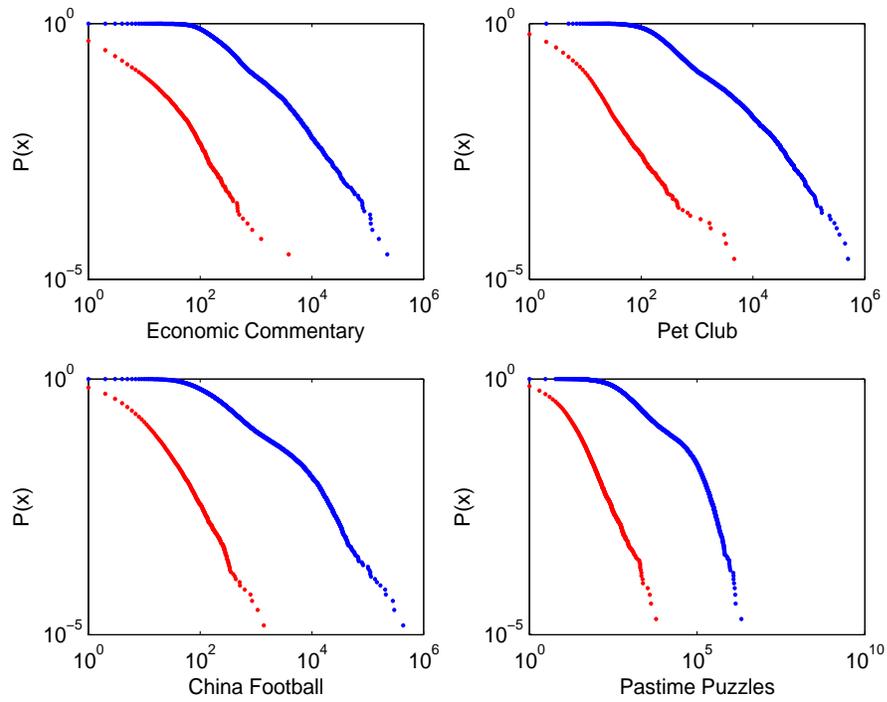}
\caption{ The cumulative distribution functions $P(x)$ and their
maximum likelihood power-law fits, for the posts of the amounts of
view and reply from $Sina$ Community Forum.The red lines stand for
the reply of cumulative distribution and the blue lines stand for
the view of cumulative distribution. } \label{power}
\end{center}
\end{figure}

$Chat Avenue$ is the current leader in free online chat rooms. Most
of users are from the USA, Canada, Australia and the United Kingdom.
And it offers a very active message forum with over 120,000 members,
an arcade, and photo gallery \cite{12}. We got the amounts of view
and reply about posts of World Events-Politics and News, Jokes,
General, and Graphics Forum.

World Events-Politics and News: From June 10, 2006 to February 23,
2009, there were 1,558 posts. Jokes: From June 10, 2006 to February
25, 2009, there were 1,583 posts. General: From June 10, 2006 to
March 3, 2009, there were 1,250 posts Graphics Forum: From June 10,
2006 to March 13, 2009, there were 1,131 posts.

According to the number of posts from $Sina$ Forum and $Chat Avenue$
Forum, we can see that the Chinese Internet Forums are more active
than foreign ones.

\begin{figure}[th]
\begin {center}
\includegraphics[height=10cm,width=0.95\textwidth]{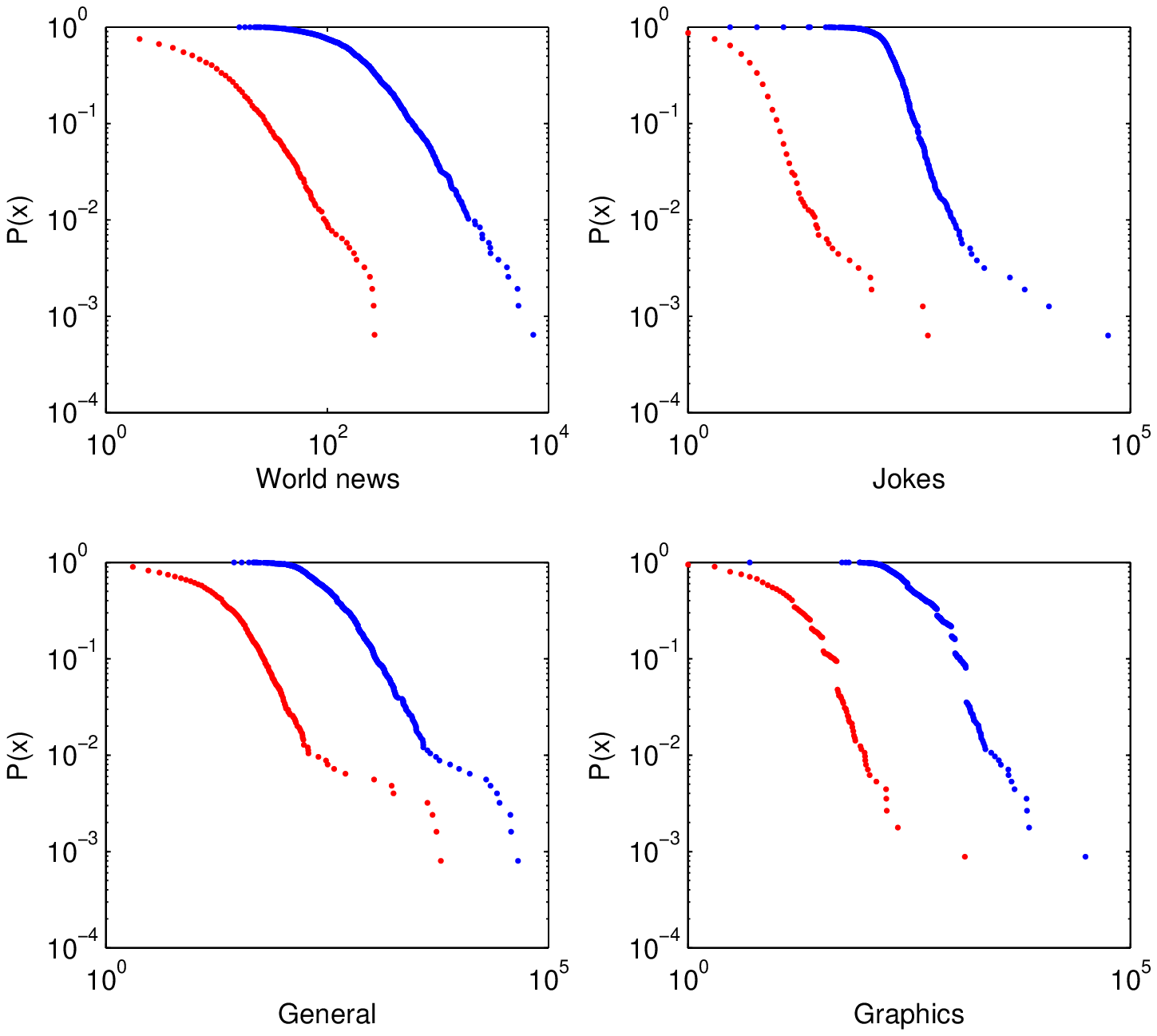}
\caption{ The cumulative distribution functions $P(x)$ and their
maximum likelihood power-law fits, for the posts of the amounts of
view and reply from $Chat Avenue$ Forum. The red lines stand for the
reply of cumulative distribution and the blue lines stand for the
view of cumulative distribution. } \label{fp}
\end{center}
\end{figure}

A power law distribution is observed when we inspect the total
number of views and replies a post receives (see FIG.\ref{power} and
FIG.\ref{fp}), indicating that the vast majority of posts generates
little interest, while a few posts are highly popular. At the same
time, the distribution of the amounts of view and reply has the
different power exponents. Mostly the power exponents of reply are
larger than that of view, although the amounts of reply are smaller
than that of view (see Table \ref{valuepower}). The power exponents
$\alpha$ range from $1.6$ to $3.5$. The values of power exponent and
$x_{min}$ are based on the method of Ref.\cite{13}.

From Table \ref{valuepower}, we also can discover that the exponents
of view from $Chat Avenue$ Forum are bigger than those of view from
$Sina$ Community Forum. But the exponent of reply is no much
difference. This shows that the viewers of posts from $Sina$
Community Forum have greater range, this may be related to the forum
rules --- more attended posts will be topped and forum also can
recommend some wonderful posts to Internet users.

\begin{table}
\caption{The values of power exponent and $x_{min}$ of different
themes }\label{valuepower}
\begin{tabular} [t]{|c|c|c|c|c|c|}
\hline Forum &Theme&\multicolumn{2}{|c|}{View}&
\multicolumn{2}{|c|}{Reply}\\\cline{3-6}
             & &$\alpha$& $x_{min}$&$\alpha$& $x_{min}$\\\hline   &Economic Commentary&2.05& 256&2.91&
              57\\\cline{2-6}
 $Sina$&Pet Club &1.94 &207&2.66& 13\\\cline{2-6}&China Football&1.87& 119&3.06 &60\\\cline{2-6}  &Pastime Puzzles&1.62
&92&2.51& 46\\\hline  &World News &2.57&382&2.81&26\\\cline{2-6}
$Chat Avenue$&Jokes &3.5&213&3.31&7\\\cline{2-6} &General
&2.52&570&2.5&31\\\cline{2-6} &Graphics &1.98&175&3.22&49\\\hline
\end{tabular}
\end{table}

Besides, we also study other themes from $Sina$ Forum, all of them
have the similar broad distribution. In the Table \ref{others}, we
list the values of power exponent and $x_{min}$ of other themes from
$Sina$ Forum. And they have the similar characteristic, namely the
values of power exponent range from $1.6$ to $3.5$, and the power
exponents of view are smaller than those of reply.

\begin{table}
\caption{The values of power exponent and $x_{min}$ of other themes
from $Sina$ Forum }\label{others}
\begin{tabular} [t]{|c|c|c|c|c|c|}
\hline Forum &Theme&\multicolumn{2}{|c|}{View}&
\multicolumn{2}{|c|}{Reply}\\\cline{3-6}
             & &$\alpha$& $x_{min}$&$\alpha$& $x_{min}$\\\hline   &Arrange Finance&2.07& 232&2.12&
             9\\\cline{2-6}
 &Education Viewpoint &1.61 &1121&1.92& 3\\\cline{2-6}&Postgraduate Communication&2.04& 2372&2.13 &5\\\cline{2-6}  &Diet
 Forum&1.92
&493&3.28& 17\\\cline{2-6} $Sina$ &08European Cup
&2.05&162&2.55&35\\\cline{2-6} &Film Television World
&1.81&1144&2.74&94\\\cline{2-6} &People Gossip
&2.12&896&2.38&38\\\cline{2-6} &The ebb and flow of Overseas
&1.93&394&2.69&47\\\cline{2-6} &Words on
World&2.09&13689&2.18&8\\\hline
\end{tabular}
\end{table}

Some questions should be put forward, for example, why the
distribution of the amounts of view and reply follow the power but
have the different power exponents, and what relationship exists
between the views and replies. We will solve these questions in the
next section.


\section{The relationship of views and replies}\label{sect3}

\begin{figure}
\begin {center}
\includegraphics[height=10cm,width=0.95\textwidth]{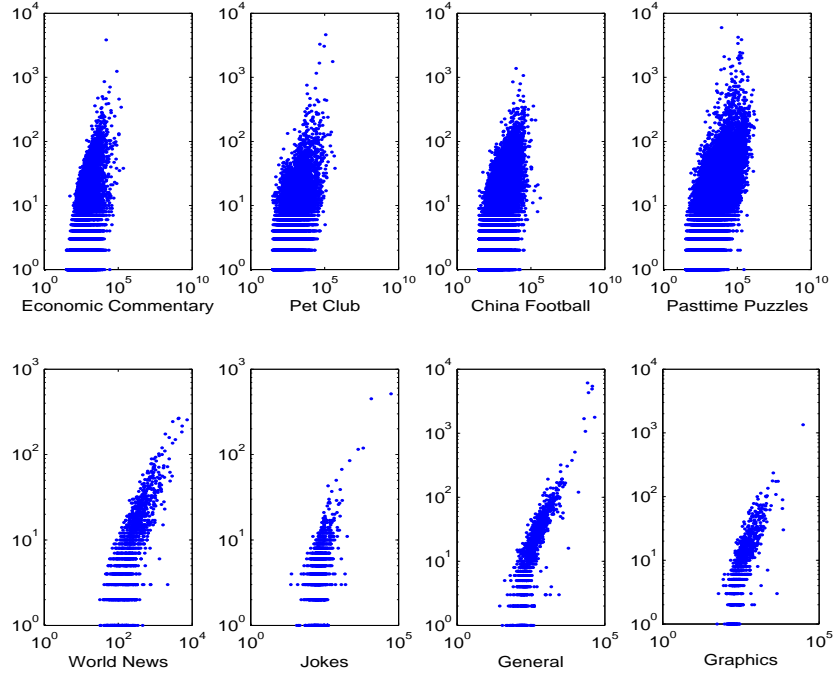}
\caption{ The relationship between the amounts of view and those of
reply is approximately linear in the log-log
coordinate.}\label{scatterplot}
\end{center}
\end{figure}

The relationship between the amounts of view and those of reply can
be observed through the FIG.\ref{scatterplot}, and it shows
non-linearity which could be described by Eq.(\ref{eq1}).
\begin{equation}
\log y=A+B\log x\label{eq1}
\end{equation}
where $y$ stands for the amounts of reply, $x$ stands for the
amounts of view. From analysis of Sect.\ref{sect2}, we know that the
amounts of view and those of reply can be described by power law
distribution. Therefore, here we suppose that Probability Density
Function of views is
\begin{equation}
f(x)=c_1 x^{-r_1}\label{eq2}
\end{equation}
The Probability Density Function of replies is
\begin{equation}
f(y)=c_2 y^{-r_2}\label{eq3}
\end{equation}
From Eq.(\ref{eq1}) and Eq.(\ref{eq3}), we can get that
 \begin{equation}
f(y)=\frac{c_1}{B}e^{\frac{A(r_1-1)}{B}}
y^{-(1+\frac{r_1-1}{B})}\label{eq4}
\end{equation}
Contrasting to Eq.(\ref{eq3}), we can find that
\begin{equation}
c_2=\frac{c_1}{B}e^{\frac{A(r_1-1)}{B}}\label{eq5}
\end{equation}
\begin{equation}
r_2=1+\frac{r_1-1}{B}\label{eq6}
\end{equation}
From Eq.(\ref{eq5}) and Eq.(\ref{eq6}) we can evaluate that
\begin{equation}
B=\frac{r_1-1}{r_2-1}
\end{equation}
\begin{equation}
A=\frac{1}{r_2-1}\left(\log B+\log c_2 -\log c_1 \right)
\end{equation}
Here $r_1$ and $r_2$ are known quantities, so we must evaluate
unknown quantities $c_1$ and $c_2$.
 We know the Probability Density
Function is
\begin{equation}
f(z)=\left\{\begin{array}{ll} cz^{-r}&\textrm{$0<z<z_{max}$}\\
0 &\textrm{others}
\end{array}\right.
\end{equation}
According to the characteristic of the Probability Density Function
\begin{equation}
\sum_{z_{min}}^{z_{max}}cz^{-r}=\frac{n}{N}
\end{equation}
We can evaluate the value of $c$.
\begin{equation}
c=\frac{n}{N}\frac{1}{\sum_{z_{min}}^{z_{max}} z^{-r}}
\end{equation}
where $n$ stands for the number of $z_{min}<z<z_{max}$, $z_{min}$
can be used from Table \ref{valuepower}, $z_{max}$ is the largest
number which can easily get from the series of views or replies. $N$
stands for the all number of $z$. Therefore, we can easily evaluate
the value of $c_1$ and $c_2$
\begin{equation}
c_1=\frac{n_1}{N}\frac{1}{\sum_{x_{min}}^{x_{max}} x^{-r_1}}
\end{equation}
\begin{equation}
c_2=\frac{n_2}{N}\frac{1}{\sum_{y_{min}}^{y_{max}} y^{-r_2}}
\end{equation}
So we can get the value of $A$
\begin{equation}
A=\frac{1}{r_2-1}\left(\log \frac{r_1-1}{r_2-1}+ \log
\frac{n_2}{n_1}+\log \frac{\sum_{x_{min}}^{x_{max}}
x^{-r_1}}{\sum_{y_{min}}^{y_{max}}y^{-r_2}} \right)
\end{equation}

According to Eq.(\ref{eq1}), we can know that the fitted line is
determined by the parameters $A$ and $B$. For the same amounts of
view, when the value of $B$ is larger, there would be more people to
reply the post. The value of $e^A$ gives the probability of the
first viewer to reply the post. The value of $e^{-A/B}$ gives the
number of people who had viewed the post before there is the first
reply to the post.

In terms of the above analysis, from  Table \ref{value}, we can know
that when there is the same amounts of view, the groups in the $Chat
Avenue$ Forum reply more than the groups in the $Sina$ Forum. Most
people in the $Sina$ Forum prefer viewing to replying the posts. The
differences between $Chat Avenue$ Forum and $Sina$ Forum come from
the different culture and social environment possibly.

\begin{table}
\caption{The values of $A$ ,$B$ and $-A/B$}\label{value}
\begin{tabular} [t]{|c|c|c|c|c|}
\hline Forum&Theme&$A$&$B$&$-A/B$\\\hline &Economic Commentary
&-2.74 &0.550 &4.98\\\cline{2-5} $Sina$&Pet Club &-3.35 &0.566&5.92
\\\cline{2-5} & China
Football&-3.44 &0.422&8.15
\\\cline{2-5}
&Pastime Puzzles&-3.70 &0.411&8.39
\\\hline
&World News &-5.17&0.967&5.35\\\cline{2-5} $Chat Avenue$& Jokes
&-7.83&1.08&7.25\\\cline{2-5}& General &-3.53&1.01&3.50\\\cline{2-5}
&Graphics &-0.456&0.441&1.03\\\hline
\end{tabular}
\end{table}

\section{Burstiness and memory}\label{sect4}

 The dynamics of a wide range of real systems display a bursty,
 characterized by short timeframes of intense activity followed
 by long times of no or reduced activity.
 In Ref.\cite{13} K.-I. Goh and  A.-L. Barab\'asi propose to characterize the bursty
 nature of real signals using orthogonal measures quantifying
 two distinct mechanisms leading to burstiness:
 the interevent time distribution (see Eq.\ref{eq12}) and the memory(see Eq.\ref{eq13}).
\begin{equation}
B=\frac{\sigma_{\tau}-m_{\tau}}{\sigma_{\tau}-m_{\tau}}
 \label{eq12}
\end{equation}
where $m_{\tau}$ and $\sigma_{\tau}$ are the mean and the standard
deviation of $P(\tau)$, respectively.
\begin{equation}
M=\frac{1}{n_{\tau}}\sum_{i=1}^{n_{\tau-1}}\frac{(\tau_{i}-m_{1})(\tau_{i+1}-m_2)}{\sigma_1\sigma_2}
  \label{eq13}
\end{equation}
where $n_{\tau}$ is the number of interevent times measured from the
signal and $m_1(m_2)$ and $\sigma_1(\sigma_2)$ are sample mean and
sample standard deviation of $\tau_i$'s ($\tau_{i+1}$'s),
respectively($i$=$1,..., n_{\tau} -1$).

The time of posting and replying can be as a time signal, then we
can easily calculate the interevent time distribution and the
memeory (see Table \ref{tab3}).

\begin{table}
\caption{The values of $ B$ and $M$}\label{tab3}
\begin{tabular} [t]{|c|c|c|c|}
\hline Forum &Theme&$B$&$M$\\\hline &People Gossip &0.8243 &0.0378
\\\cline{2-4} $Sina$& The ebb and flow of Overseas &0.7856&0.0457
\\\cline{2-4} &Words on World&0.7296&0.01108
\\\hline
&World News&0.8897 &0.0057
\\\cline{2-4}
$ChatAvenue$ &Jokes &0.8337&0.0057\\\cline{2-4}
 &Graphics &0.8258&0.0348\\\hline
\end{tabular}
\end{table}

In Table \ref{tab3}, the themes of People Gossip, The ebb and flow
of Overseas and Words on World belong to $Sina$ Community Forum, and
the themes of World News, Jokes and Graphics come from $Chat Avenue$
Forum. From the Table \ref{tab3}, we can discover that all the
themes of the values $B$ and $M$ have no much differences which will
further explain that the replying behaviors of $Sina$ Community
Forum and $Chat Avenue$ Forum are similarity. All of them can
illustrate that the replying time series are bursty signal, in other
words, the front people who reply the post can not affect the behind
ones, they would reply a post depending on their interests.

\section{Conclusion and discussion}

In this paper we discovered that the amounts of view and reply on
$BBS$ forum appear the power law distributions, the power exponents
range from $1.6$ to $3.5$. Moreover, mostly the power exponents of
reply are bigger than that of view. Furthermore, we found that the
amounts of view and those of reply has the nonlinear relationship.
According to the relationship, we presented two parameters --- the
slope and intercept of the log-log straight line which can evaluate
human's interests on replying or viewing. With the above
measurements and statistics, we find some differences between $Sina$
Community Forum and $Chat Avenue$ Forum, which may be caused by
cultural, social environments, and forum rules. At last, we analyze
the burstiness and memory in replying time series. It could be found
that all the time series have similar qualitative properties. The
$Sina$ Community Forum and $Chat Avenue$ Forum have no much
differences in replying time series.

\section*{Acknowledgement}
This paper is supported by NSFC under the grant Nos.
70771011,70471080.

\end{document}